\documentclass[a4paper]{jpconf}
\usepackage{multicol}
\usepackage{graphicx}
\usepackage{amsfonts}
\begin{document}

\title{Energy spectrum of strongly correlated particles in quantum dots}

\author{K. Balzer, C.N\"olle, M. Bonitz and A. Filinov}

\address{Christian-Albrechts-Universit\"at Kiel, Institut f\"ur Theoretische Physik und Astrophysik, Leibnizstrasse 15, 24098 Kiel, Germany}

\begin{abstract}
The ground state and the excitation spectrum of strongly correlated electrons in quantum dots are investigated. An analytical solution is constructed by exact diagonalization of the  Hamiltonian in terms of the $N$-particle eigenmodes.
\end{abstract}

\section{Introduction}
Spatially confined charged particles are attracting increasing interest. In particular,
quantum confined semiconductor structures allow to design materials with novel
properties, see e.g. Ref. \cite{ashoori} for an overview. Of special interest are
quantum dots which allow to confine charged particles in all three space dimensions,
e.g. \cite{reimann}. Besides their
application in nanoscale transport and optics, quantum dots allow to create strongly
correlated many-electron states including Fermi liquid and Wigner crystal behaviour
\cite{afilinov-etal.01prl} which are of fundamental interest in many fields. While the ground state of these systems
can be studied theoretically from first principles, e.g. using quantum Monte Carlo
methods, these methods fail to deliver the energy spectrum. Therefore,
various analytical concepts have been considered which allow to treat small systems of confined electrons.
The problem of the energy spectrum of two electrons in a harmonic trap has been solved in Refs.
\cite{peeters01,loz03}.

Here, we extend the analytical approach of Ref. \cite{loz03} to the
general case of $N$ strongly correlated particles (e.g. electrons). We demonstrate
in Sec. $1$ that, in the strong-coupling limit, the $N$-particle problem can
be reduced to a superposition of $(N\cdot d)$ uncoupled harmonic oscillators.
These results are illustrated for two electrons in 1$d$ and 2$d$ in Sec. $3$ and $4$
where we demonstrate how to analytically construct the ground state wave function
and energy spectrum in the whole range from strong to moderate coupling.

\section{Energy spectrum of $N$ strongly correlated electrons}
The Hamiltonian of $N$ Coulomb-interacting electrons of effective mass $m$ trapped in a $d$-dimensional harmonic potential is given by
\begin{equation}
\label{eq:1}
	\hat{H} = T(\hat{\vec{p}}_1,\ldots,\hat{\vec{p}}_N)+V(\vec{r}_1,\ldots,\vec{r}_N)
	= \sum^{N}_{i=1}{\frac{\hat{\vec{p}}_{i}^{\,2}}{2 m}}+\left(\sum^{N}_{i=1}{\frac{m}{2}\omega^2 \vec{r}_{i}^{\,2}}+\sum^{N}_{i<j}{\frac{\alpha}{|\vec{r}_i-\vec{r}_j|}}\right)\;,
\end{equation}
where $\omega$ is the confinement frequency, and \mbox{$\alpha=\frac{e^2}{4 \pi \epsilon_0 \epsilon_b}$} is the Coulomb interaction strength, with $\epsilon_b$ being the background dielectric constant.

The degree of electron-electron correlations is characterized by the dimensionless coupling parameter $Q$ \cite{loz04}, which is the ratio of the characteristic Coulomb energy $E_B$ to the confinement energy $\hbar\omega$.
\begin{equation}
\label{eq:2}
	Q=\left(\frac{E_B}{\hbar\omega}\right)^{\frac{1}{3}}=\left(\frac{a_0}{2 a_B}\right)^{\frac{2}{3}}=\frac{1}{\hbar}\left({\frac{\alpha^2 m}{\omega}}\right)^{\frac{1}{3}}\,,
\end{equation}
where $E_B=\frac{\alpha}{2 a_B}=\frac{m \alpha^2}{2 \hbar^2}$, $a_0=\sqrt{\frac{\hbar}{m\omega}}$ is the characteristic confinement length, and $a_B$ denotes the effective Bohr radius. In the strong-coupling limit, $Q\gg1$, the electrons are spatially localized around their classical equilibrium positions $\vec{r}_0=(\vec{r}_{10},\ldots,\vec{r}_{N0})$, and the confinement potential $V(\vec{r})$ can be well approximated harmonically around $\vec{r}_0$.
\begin{equation}
\label{eq:3}
	V(\vec{r})= V(\vec{r}_0)+\frac{1}{2!}(\vec{r}-\vec{r}_0)^T A \;(\vec{r}-\vec{r}_0)+\delta V\;,
\end{equation}
where $A_{ij}=\frac{\partial^2 V(\vec{r}_0)}{\partial r_i \partial r_j}$ with \mbox{$\vec{r}=(r_1,\ldots,r_{N\cdot d})$} is the $N\cdot d$-dimensional Hesse matrix, and the anharmonic correction $\delta V$ is neglected, it will be restored in Sec. $5$. Further, since matrix $A$ is symmetric there exist square $N\cdot d$-dimensional matrices $B$ and $U$, such that $A = U^T\,B\,U$, where $B$ is diagonal and $U$ is unitary. Introducing (\ref{eq:3}) into (\ref{eq:1}) yields
\begin{eqnarray}
\label{eq:5}
	\hat{H} - V(\vec{r}_0)&=& \frac{1}{2 m} (U\,\hat{\vec{p}}\,)^{T} (U\,\hat{\vec{p}}\,)+\,\frac{1}{2}(U\,(\vec{r}-\vec{r}_0))^T\,B\,(U (\vec{r}-\vec{r}_0)) \\
	&=& \frac{1}{2 m} (\hat{\vec{P}})^{T} (\hat{\vec{P}})\,+\,\frac{1}{2}(\vec{R})^T\,B\,(\vec{R})\nonumber\;,
\end{eqnarray}
where \mbox{$\hat{\vec{p}}=(\hat{p}_{1},\ldots,\hat{p}_{N\cdot d})$}, and the new momenta and space coordinates, $\hat{\vec{P}}$ and $\vec{R}$, are defined by
\begin{equation}
\label{eq:6}
	\hat{\vec{P}} = U\,\hat{\vec{p}}\;,\hspace{2pc}\vec{R} = U\,(\vec{r}-\vec{r}_0)\;.
\end{equation}
With the diagonal elements $B_{ii} = m\,\Omega^{2}_{i}\,$, Eq. (\ref{eq:5}) simplifies to
\begin{equation}
\label{eq:8}
	\hat{H}\,-\,V(\vec{r}_0)= \sum^{N\cdot d}_{i=1}{\left(\frac{\hat{P}^{2}_{i}}{2 m}+\frac{m}{2}\Omega^{2}_{i} R^{2}_{i}\right)}\;,
\end{equation}
and the $N$-particle problem has been reduced to a superposition of $(N\cdot d)$ uncoupled harmonic oscillator modes with the eigenfrequencies
\begin{equation}
\label{eq:9}
\Omega_i(\omega)=\sqrt{\frac{B_{ii}}{m}}\;.
\end{equation}
This description in terms of normal modes arises in the harmonic approximation of the potential around the classical equilibrium position $\vec{r}_0$, which is exact for strong electron-electron correlations, $Q\rightarrow \infty$. The eigenfunctions of the $N$-particle Hamiltonian of Eq. (\ref{eq:8}), $\Psi_{\vec{n}}(\vec{R})$, can be written as a product of $(N\cdot d)$ one-dimensional oscillator eigenfunctions
\begin{equation}
\label{eq:10}
	\Psi_{\vec{n}}(\vec{R})=\prod^{N\cdot d}_{i=1}{\psi_{n_i}(R_i})\;,\hspace{2pc}\psi_{n_i}(R_i) = c_{n_i} \exp{\left(-\frac{{\tilde{R}}^{2}_{i}}{2}\right)}\,H_{n_i}({\tilde{R}}_i)\;,
\end{equation}
\begin{equation}
\label{eq:add10}
	c_{n_i} = \frac{1}{\sqrt{2^{n_i} n_i!}}\left(\frac{\Omega_i}{\pi\,\omega}\right)^{\frac{1}{4}}\;,\hspace{2pc}R_i=\sqrt{\frac{\hbar}{m\,\Omega_i}}{\tilde{R}}_i\;.
\end{equation}
where $\vec{n}=(n_1,\ldots,n_{N\cdot d})$, $n_i=0,1,2,\ldots\;$, and $H_{n_i}({\tilde{R}}_i)$ are the Hermite polynomials. The corresponding excitation spectrum around the classical ground state $V(\vec{r}_0)$ is given by
\begin{equation}
\label{eq:11}
	E_{\vec{n}} = V(\vec{r}_0)+\sum^{N\cdot d}_{i=1}{\hbar\sqrt{\frac{B_{ii}}{m}}\left(n_i+\frac{1}{2}\right)}\;.
\end{equation}
In the following two Sections, our general approach is illustrated for $N=2$ strongly correlated electrons in 1$d$ and 2$d$.

\section{Two electrons in a 1$d$ harmonic trap}
For two electrons in a one-dimensional harmonic confinement
\begin{equation}
\label{eq:12}
	\hat{H} = T(\hat{p}_1,\hat{p}_2)+V(x_1,x_2) = \sum^{2}_{i=1}{\frac{\hat{p}_{i}^2}{2 m}}+\left(\sum^{2}_{i=1}{\frac{m}{2}\omega^2 x_{i}^2}+\sum^{2}_{i<j}{\frac{\alpha}{|x_i-x_j|}}\right)\;,
\end{equation}
\begin{equation}
\label{eq:13}
	V(\vec{x})\simeq V(\vec{x}_0)+\frac{1}{2!}(\vec{x}-\vec{x}_0)^T A \;(\vec{x}-\vec{x}_0)\;,
\end{equation}
where $\vec{x}=(x_1,x_2)$, and $\vec{x}_0=(x_{10},x_{20})$ is the classical equilibrium position of the electons, which follows from $\nabla\; V(\vec{x}_0)=0$ :
\begin{equation}
\label{eq:14}
	\vec{x}_0=\frac{1}{2}(-x_0,x_0)\;,\hspace{2pc}x_0=\left(\frac{2\alpha}{m\omega^2}\right)^{1/3}\;.
\end{equation}
The equilibrium potential energy is
\begin{equation}
\label{eq:15}
	V(\vec{x}_0)=\frac{3}{2}\left(\frac{m}{2}\alpha^2\omega^2\right)^{1/3}=\hbar\,\omega\frac{3}{2^{4/3}}Q\;,
\end{equation}
and, using (\ref{eq:14}), the matrix $A$ becomes
\begin{equation}
\label{eq:17}
	A=\;
	U^T \underbrace{
	\left(\begin{array}{cc}
	{m \omega^2} & {0}\\
	 {0}	& {3 m \omega^2}
	\end{array}\right)}_{=B}
	\;U\;,\hspace{2pc}
	U=\frac{1}{\sqrt{2}}
	\left(\begin{array}{cc}
	1 & 1\\
	-1 & 1
	\end{array}\right)\;.
\end{equation}
The unitary matrix $U$ (\ref{eq:17}) defines the canonical transformation to the new variables $(\tilde{x}_1,\tilde{p}_1)$ and $(\tilde{x}_2,\tilde{p}_2)$,
\begin{equation}
\label{eq:19}
	\left(\begin{array}{c}
	{\tilde{x}_1} \\
	 {\tilde{x}_2}
	\end{array}\right)\;
	=U
	\left(\begin{array}{c}
	{{x}_1+\frac{x_0}{2}} \\
	 {{x}_2-\frac{x_0}{2}}
	\end{array}\right)\;
	=\frac{1}{\sqrt{2}}
	\left(\begin{array}{c}
	{x_1+x_2}\\
	 x_2-x_1-x_0
	\end{array}\right)\;,
\end{equation}
\begin{equation}
\label{eq:20}
	\left(\begin{array}{c}
	{\tilde{p}_1} \\
	 {\tilde{p}_2}
	\end{array}\right)\;
	=U
	\left(\begin{array}{c}
	{{p}_1} \\
	 {{p}_2}
	\end{array}\right)\;
	=\frac{1}{\sqrt{2}}
	\left(\begin{array}{c}
	{p_1+p_2}\\
	{p_2-p_1}
	\end{array}\right)\;.
\end{equation}
In the harmonic approximation ($\delta V=0$), the electron dynamics then can be described as a superposition of $N\cdot d=2$ uncoupled harmonic oscillator modes
\begin{equation}
\label{eq:21}
	\hat{H}-V(\vec{x}_0)=\sum^{2}_{i=1}{\left(\frac{{\tilde{p}_i}^2}{2 m}+\frac{m}{2} {\Omega^{2}_{i}} \;{\tilde{x}_i}^2\right)}\;,\hspace{2pc}\Omega_1=\omega\;,\hspace{2pc}\Omega_2=\sqrt{3}\,\omega\;,
\end{equation}
The first mode of frequency $\omega$ corresponds to the oscillation of the electron center-of-mass, whereas the second (breathing) mode, see Ref. \cite{peeters01}, of frequency $\sqrt{3}\,\omega$ describes anti-phase oscillations of the two electrons. Replacing $\tilde{p}_i$ in Eq. (\ref{eq:21}) by the momentum operator $-i \hbar \nabla_{\tilde{x}_i}$ and introducing dimensionless coordinates $\bar{x}_1$ and $\bar{x}_2$ by (cf. Eq. (\ref{eq:2})) $\tilde{x}_i=a_0\;\bar{x}_i$\;, yields the Hamiltonian
\begin{equation}
\label{eq:23}	\frac{\hat{H}-V(\vec{x}_0)}{\hbar\omega}=\sum^{2}_{i=1}{\left(-\frac{1}{2}\nabla^{2}_{\tilde{x}_i}+\frac{\eta^{2}_{i}}{2}{\tilde{x}_i}^2\right)}\;,\hspace{2pc}\eta_i={\Omega_i}/{\omega}\;.
\end{equation}
The eigenfunctions of Eq. (\ref{eq:23}), $\Psi_{\vec{n}}(\bar{x}_1,\bar{x}_2)$, are of the form of Eq. (\ref{eq:10}) with $\vec{n}=(n_1,n_2)$, $n_i=0,1,2,\ldots\;$, i.e. $\Psi_{\vec{n}}(\bar{x}_1,\bar{x}_2)=\psi_{n_1}(\bar{x}_1)\cdot\psi_{n_2}(\bar{x}_2)$, where
\begin{equation}
\label{eq:25}
	\psi_{n_i}(\bar{x}_i)=c_{n_i}\exp{\left(-\frac{{\eta_i\,\bar{x}_i}^2}{2}\right)} H_{n_i}(\sqrt{\eta_i}\,\bar{x}_i)\;,\hspace{2pc}	c_{n_i}=\frac{1}{\sqrt{2^{n_i} n_{i}!}}\left(\frac{\eta_i}{\pi}\right)^{\frac{1}{4}}\;.
\end{equation}
The two-particle excitation spectrum is then given by
\begin{equation}
\label{eq:27}
E_{n_1,n_2}=V(\vec{x}_0)+\epsilon^{(1)}_{n_1}+\epsilon^{(2)}_{n_2}\;,\hspace{2pc}	\epsilon^{(i)}_{n}=\hbar\Omega_i\left(n+\frac{1}{2}\right)\;.
\end{equation}

\begin{figure}[t]
\hspace{1pc}%
\includegraphics[width=16pc]{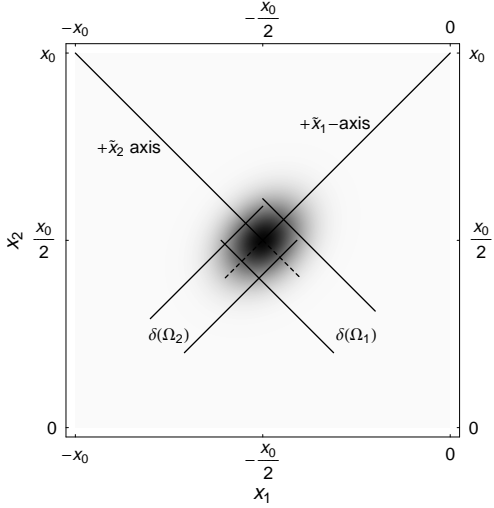}\hspace{3pc}%
\begin{minipage}[b]{18pc}
\caption{
Ground state correlation function $K_{0,0}(x_1,x_2)$ for coupling parameter $Q=50$ \mbox{($a_0\approx700\,a_B$)}. In $\tilde{x}_1$- and $\tilde{x}_2$-direction one recognizes the different widths of the peak, $\delta(\Omega_i)$.}
\end{minipage}
\end{figure}

The properties of the ground state wave function, $\Psi_{(0,0)}(\bar{x}_1,\bar{x}_2)$, can be illustrated using the two-particle correlation function in real-space
\begin{equation}
\label{eq:28}
	K_{\vec{n}}(x,x^{\prime})=\int{dx_1}\int{dx_2}|\Psi_{n_1,n_2}(\bar{x}_1,\bar{x}_2)|^2
	\cdot \left[\hat\rho(x)\hat\rho(x^{\prime})-\delta(x-x^{\prime})\hat\rho(x)\right]\;,
\end{equation}
where $\hat{\rho}(x^{(\prime)})=\sum^{2}_{i=1}{\delta(x^{(\prime)}-x_i)}$. Eq. (\ref{eq:28}) simplifies to \cite{peeters01}
\begin{equation}
\label{eq:29}
		K_{\vec{n}}(x_1,x_2)=2 \; |\psi_{n_1}(\bar{x}_1)|^2 \; |\psi_{n_2}(\bar{x}_2)|^2\;,
\end{equation}
where from Eq. (\ref{eq:19}) it is $\bar{x}_1=\frac{x_1+x_2}{\sqrt{2}\,a_0}$ and $\bar{x}_2=\frac{x_2-x_1-x_0}{\sqrt{2}\,a_0}$. The density plot of $K_{0,0}(x_1,x_2)$ is presented in Fig. $1$. The ground state correlation function shows a Gaussian-like peak around the position $(x_1,x_2)=(-\frac{x_0}{2},\frac{x_0}{2})$, which has two different characteristic widths $\delta(\Omega_i)$ in directions $\tilde{x}_1$ and $\tilde{x}_2$ of the corresponding eigenmodes.

Next, consider the ground state electron density $\rho_{(0,0)}(x)$ defined by \cite{peeters01}
\begin{equation}
\label{eq:30}
		\rho_{\vec{n}}(x)=\int{dx_1\int{dx_2 |\Psi_{\vec{n}}(\tilde{x}_1,\tilde{x}_2)|^2\hat\rho(x)}}\;,
\end{equation}
where for $n_1=n_2=0$ we obtain
\begin{equation}
\label{eq:31}
\rho_{0,0}(x)=\frac{1}{\rho_0}\left(e^{-\frac{1}{\kappa}\left(1+\frac{2x}{x_0}\right)^2}+e^{-\frac{1}{\kappa}\left(1-\frac{2x}{x_0}\right)^2}\right)\;,\hspace{2pc}\kappa=\frac{2\;(1+\sqrt{3})\;\hbar}{\sqrt{3} \;m\; \omega \; x^{2}_{0}}\;,
\end{equation}
with $\kappa$ being the dimensionless variance. The factor $1/{\rho_0}$ ensures that the electron density is normalized to the particle number, $\int{\rho_{0,0}(x) dx}=2$. $\rho_{0,0}$ is shown in Fig. $2$ for parameters $Q$ in the moderate and strong coupling limit. The variance $\kappa$ determines the width of the peaks around the equilibrium positions $-\frac{x_0}{2}$ and $\frac{x_0}{2}$. Using (\ref{eq:14}) we obtain $\kappa\propto\omega^{1/3}\propto \frac{1}{Q}$. In the strong-coupling limit, it is $\kappa\ll 1$ corresponding to strong electron localization. For $Q\,\widetilde{<}\,1$, strong electron overlap is observed, which, however, is beyond the harmonic approximation of Eq. (\ref{eq:3}), and one has to account for anharmonic corrections $\delta V$, see Sec. 5.

\begin{figure}[t]
\includegraphics[width=18pc]{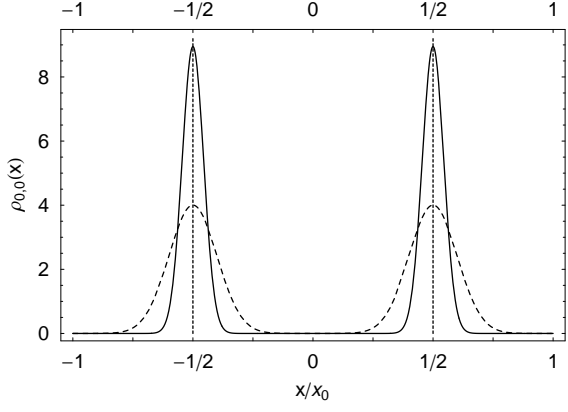}\hspace{2pc}%
\begin{minipage}[b]{18pc}
\caption{
Ground state electron density $\rho_{0,0}(x)$ for $Q=25$ (dashed curve) and $Q=125$ (solid curve).}
\end{minipage}
\end{figure}

\section{Extension to two electrons in 2$d$}
Here, for the case of two harmonically confined electrons in 2$d$ we analyze the correlation function $K(\vec{r},\vec{r}^{\,\prime})$. Using the notation $\vec{r}=(\vec{r}_1,\vec{r}_2)=(x_1,y_1,x_2,y_2)$, the Hamiltonian reads
\begin{equation}
\label{eq:34}
	\hat{H}=\frac{1}{2 m}(\vec{p})^{\,T}(\vec{p})+\frac{m}{2}\omega^2 (\vec{r})^{\,T}(\vec{r})+\frac{\alpha}{|\vec{r}_1-\vec{r}_2|}\;,
\end{equation}
where $\vec{p}=(\vec{p}_1,\vec{p}_2)$. The classical equilibrium distance of the electrons is given by  
\begin{equation}
\label{eq:35}
r_0=\left(\frac{\alpha}{4 m \omega^2}\right)^{\frac{1}{3}}=2^{\frac{1}{3}}a_0\,\sqrt{Q}\;.
\end{equation}
In the strong-coupling limit, the total potential can be expanded as in Eq. (\ref{eq:3}) with $\delta V=0$ resulting in
\begin{equation}
\label{eq:36}
	U=\frac{1}{\sqrt{2}}\left(\begin{array}{cccc}
	0&-1&0&1\\
	0&1&0&1\\
	1&0&1&0\\
	-1&0&1&0
	\end{array}\right)\;,\hspace{2pc}
	B=m\,\omega^2\,\left(\begin{array}{cccc}
	3&0&0&0\\
	0&1&0&0\\
	0&0&1&0\\
	0&0&0&0
	\end{array}\right)\;.
\end{equation}
With unitary transformations of Eq. (\ref{eq:6}) the two-particle problem, Eq. (\ref{eq:34}), is reduced to four uncoupled harmonic oscillators corresponding to two centre-of-mass modes $\Omega_{2,3}=\omega$, one breathing mode $\Omega_1=\sqrt{3}\,\omega$ and the trivial rotation of the whole system with $\Omega_4=0$.

The ground state correlation function (\ref{eq:28}) is evaluated to
\begin{equation}
\label{eq:37}		
	K_0(\vec{r},\vec{r}^{\,\prime})
=\frac{k_0}{\sqrt{r}}\,\exp\left(-(D^{2}_{x}+D^{2}_{y})-\frac{\sqrt{3}}{4}\left(r-2^{\frac{1}{3}}\sqrt{Q}\right)\right)	\end{equation}
where $D_x=\frac{x_1+x_2}{2 a_0}$, $D_y=\frac{y_1+y_2}{2 a_0}$, $r=\frac{1}{a_0}|\vec{r}_1-\vec{r}_2|$ and $k_0$ denotes a normalization constant.

In Fig. 3, $K_0$ is shown for the strong coupling case $Q=250$ and one electron being fixed in three different positions. Fig. 3 c) shows the correlation function when one electron is situated in its classical equilibrium position, e.g. $\vec{r}_{10}=(\frac{r_0}{2\,a_0},0)$. The density of the second electron now shows a Gaussian-like peak around its respective equilibrium position $\vec{r}_{20}=-\vec{r}_{10}$. As in Sec. 3 (Fig. 1) one recognizes characteristic widths in different directions corresponding to the different eigenmodes.

\begin{figure}[b]
\begin{minipage}{40pc}
\includegraphics[width=12.45pc]{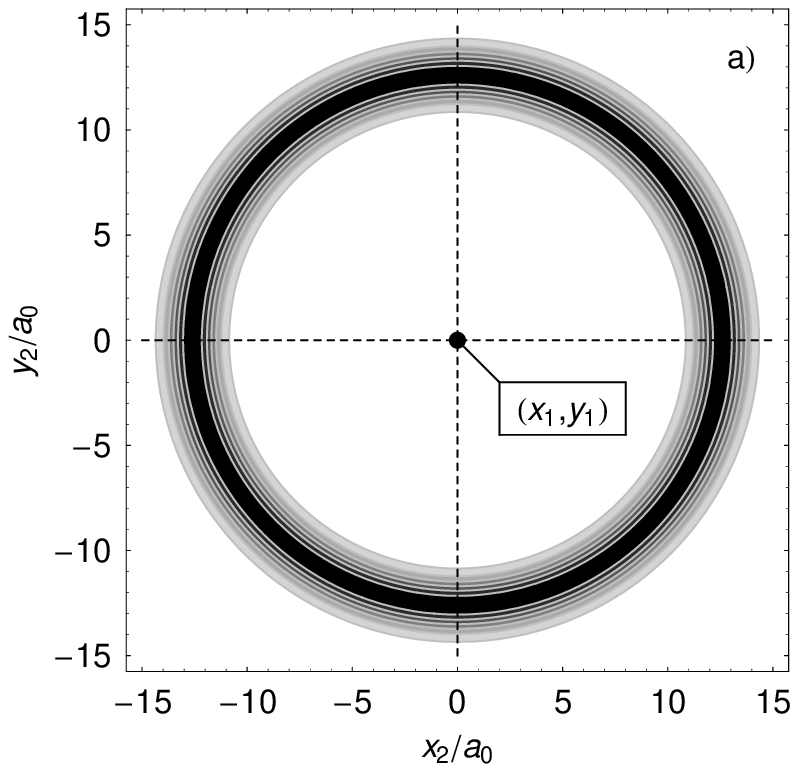}
\includegraphics[width=12.45pc]{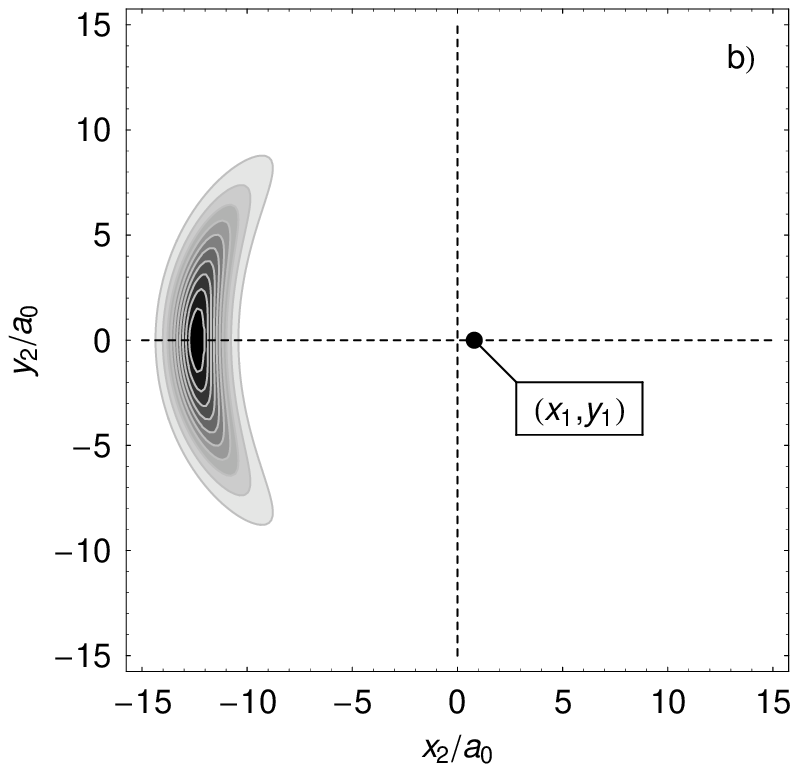}
\includegraphics[width=12.45pc]{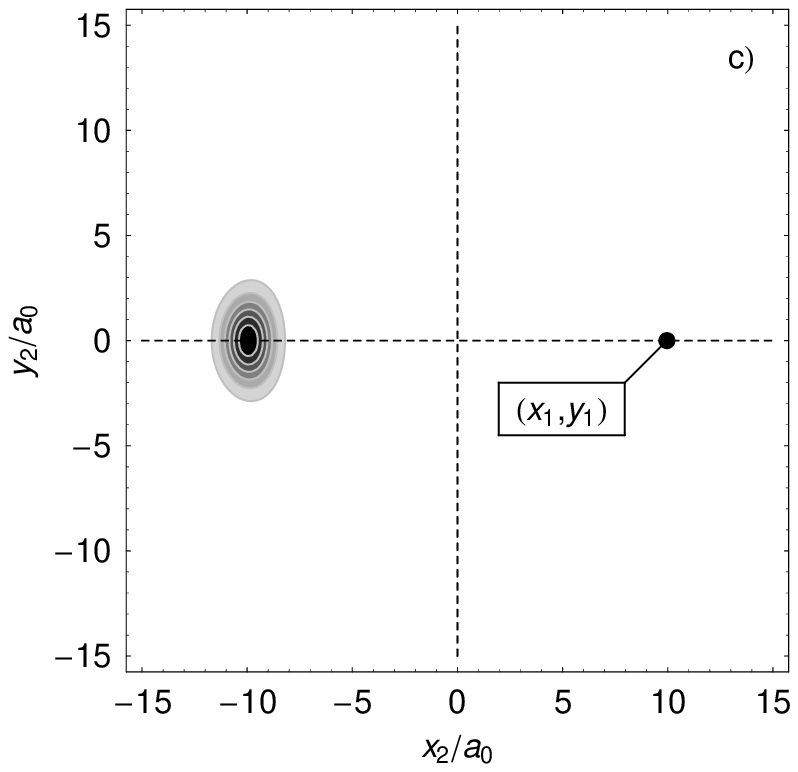}
\end{minipage}
\caption{Ground state correlation function $K_0(\vec{r},\vec{r}^{\,\prime})$ for $Q=250$. One electron (dot) is fixed in different positions $\vec{r}_1=(x_1,y_1)$ and the density of the other electron is shown, \cite{peeters01}. Lengths are displayed in units of $a_0$. Configurations of single figures: a) $\vec{r}_1=(0,0)$, b) $\vec{r}_1=(0.8,0)$ and c) $\vec{r}_1=\left(\frac{r_0}{2\,a_0},0\right)=(9.96,0)$, corresponding to the classical equilibrium position.}
\end{figure}

\section{Anharmonic corrections}
So far, we confined ourselves to the strong-coupling limit, $Q\gg1$, where $\delta V=0$ is adequate. In contrast, at weak coupling, in Eq. (\ref{eq:3}) the anharmonic corrections $\delta V$ have to be taken into account. Since the derived eigenfunctions $\Psi_{\vec{n}}(\vec{R})=:|\vec{n}\rangle$ (\ref{eq:10}) of the harmonic approximation form a complete basis, the higher order terms $\delta V$ can be expanded in terms of these eigenstates. Here, we limit ourselves to spinless particles in order to demonstrate the general scheme. In the following, the ground state and the excitation spectrum for $\delta V\neq0$ is derived for the one-dimensional case of Sec. 3. Using the transformations (\ref{eq:19}), we introduce new coordinates $\bar{x}_{1,2}$ given by $x_{1,2}=\mp \frac{x_0}{2}+\frac{a_0}{\sqrt{2}}(\bar{x}_1\mp\bar{x}_2)$, in terms of which the total potential energy has the form
\begin{equation}
\label{eq:40}
	V(\vec{\bar{x}})=V(\bar{x}_1)+V(\bar{x}_2)\\
	=\frac{\hbar\omega}{2}\bar{x}^{2}_1+\frac{m \omega^2}{4}(x_0+\xi \bar{x}_2)^2+\frac{\alpha}{|x_0+\xi \bar{x}_2|}\;,
\end{equation}
where $\vec{\bar{x}}=(\bar{x}_1,\bar{x}_2)$ and $\xi=\sqrt{2}\,a_0$. Obviously, anharmonic effects arise exclusively from $V(\bar{x}_2)$. The expansion of $V(\bar{x}_2)$ around $\bar{x}_2=0$ yields
\begin{equation}
\label{eq:41}
	V(\bar{x}_2)={V(\vec{x}_0)+\frac{3}{2}\hbar\omega \bar{x}^2_{2}}+\underbrace{\sum_{i\geq3}{\frac{\alpha(-\xi \bar{x}_2)^i}{x^{i+1}_{0}}}}_{=\,\delta V}\;.
\end{equation}
Expanding $\delta V$ in terms of Hermite polynomials $H_{k}(\bar{x}_2)$, the sum over $i$ in Eq. (\ref{eq:41}) can be rewritten as
\begin{equation}
\label{eq:42}
	\delta V=\sum_{k\geq0}{\nu_k(Q) \; H_{k}(\bar{x}_2)}\;,
\end{equation}
where the expansion coefficients $\nu_k(Q)$ are easily obtained by comparison of coefficients, and the $Q$-dependence arises from the $Q$-dependence of $x_0$, see Eq. (\ref{eq:14}). The Hamiltonian corresponding to coordinate $\bar{x}_2$ is then given by
\begin{equation}
\label{eq:43}
	\frac{\hat{H}(\bar{x}_2)}{\hbar\omega}=-\frac{1}{2}\nabla^{2}_{\bar{x}_2}+\frac{3}{2} \bar{x}^2_{2}\,+\sum_{k\geq0}{\frac{\nu_k(Q)}{\hbar\omega} \; H_{k}(\bar{x}_2)}\;,
\end{equation}
where we write $\hat{H}=\hat{H}(\bar{x}_1)+\hat{H}(\bar{x}_2) + V(\vec{x}_0)$. To numerically determine the ground state energy and relative wave function, $\Phi_0(\bar{x}_2)$, i.e.  the eigenfunction of $\hat{H}(\bar{x}_2)$ corresponding to \mbox{$\hat{H}(\bar{x}_2)\,\Phi_0(\bar{x}_2)=\varepsilon\;\Phi_0(\bar{x}_2)$}, we limit the summation in Eq. (\ref{eq:43}) to finite $k_{\max}\geq 3$. Then the ansatz reads
\begin{equation}
\label{eq:44}
	\Phi_0(\bar{x}_2)=\sum^{m}_{n=0}{C_n(Q) \cdot |n\rangle}\;,
\end{equation}
where \mbox{$|n\rangle=\psi_n(\bar{x}_2)$} are the basis eigenstates, see Eq. (\ref{eq:25}). Using the explicit form of $\hat{H}(\bar{x}_2)$, the coefficients follow from multiplication with $\langle n^{\prime}|$ :
\begin{equation}
\label{eq:46}
  \sum^{m}_{n=0}
  {C_n(Q)\,\left(\epsilon^{(2)}_{n}\delta_{n^\prime n}+\langle n^\prime | \delta V|n\rangle\right)}
  =
  \varepsilon\;
  {C_{n^{\prime}}(Q)}
  \,,
\end{equation}
where $\epsilon^{(2)}_{n}$ is given by Eq. (\ref{eq:27}). The matrix elements are computed as
\begin{equation}
\label{eq:47}
  \langle{n^\prime|\delta V|n}\rangle=\sum^{k_{\max}}_{k\geq0}{\nu_k(Q)\,\langle n^\prime|H_{k}(\bar{x}_2)|n\rangle}=\sum^{k_{\max}}_{k\geq0}{\tilde{\nu}_k(Q)\,\langle \tilde{n}^\prime|H_{k}(y)| \tilde{n}\rangle}\;.
\end{equation}
where $\sqrt{\eta}\,\bar{x}_2=y$, $\eta=\frac{\Omega_1}{\omega}=\sqrt{3}$, and $| \tilde{n}^\prime\rangle=\psi_{n^\prime}(\bar{x}_2/\sqrt{\eta})$. The integral $\langle \tilde{n}^\prime|H_{k}(y)|\tilde{n}\rangle$ is evaluated exactly with $s=\frac{1}{2}(n^\prime+k+n)$ and $c_n$ given by Eq. (\ref{eq:25}):
\begin{eqnarray}
\label{eq:49}
  \langle \tilde{n}^\prime|H_{k}(y)|\tilde{n}\rangle &=& \sqrt{\eta}\;c_{n^\prime} c_n \int{dy\,\exp(-y^2)\,H_{n^\prime}(y)\,H_{k}(y)\,H_{n}(y) }\nonumber\\
  &=& \sqrt{\eta}\;c_{n^\prime} c_n 
  \left\{
  \begin{array}{cc}
  \frac{2^s\,n^\prime!\,k!\,n!\,\sqrt{\pi}}{(s-n^{\prime})!\,(s-k)!\,(s-n)!} &,\; s\in\mathbb{Z}\hspace{7mm}\\[3mm]
  0 &,\; \textnormal{\itshape otherwise}
  \end{array}
  \right.\;.\nonumber
\end{eqnarray}

Below, we present results for the ground state wave function and the excitation spectrum for $k_{\max}=6$ and different coupling parameters $Q\,\widetilde{<}\,1$. The matrix eigenvalue equation (\ref{eq:46}) has been solved numerically with variable upper summation limit $m$ yielding a fixed precision of the energy eigenvalues, $\Delta\,\varepsilon\approx10^{-5}\,\hbar\omega$. The expression for $\delta V$ in Eq. (\ref{eq:41}) in terms of $y$ and $Q$, and the coefficients $\tilde{\nu}_k(Q)$ of Eq. (\ref{eq:47}) are given in the Appendix. In the following figures, the results for different values of the coupling parameter are shown. As one sees in the coefficient spectra of Fig. 4 to 6, mainly the even states contribute in the expansion of Eq. (\ref{eq:44}). Particularly, for $Q<0.1$, even coefficients are several orders of magnitude larger than the odd ones. Also, the spectrum gets broader for decreased coupling, and the values drop off approximately exponentially with $n$. !
 This property is marked by the straight lines in Fig. 4-6, corresponding to exponential fits according to

\begin{figure}[t]
\includegraphics[width=18pc]{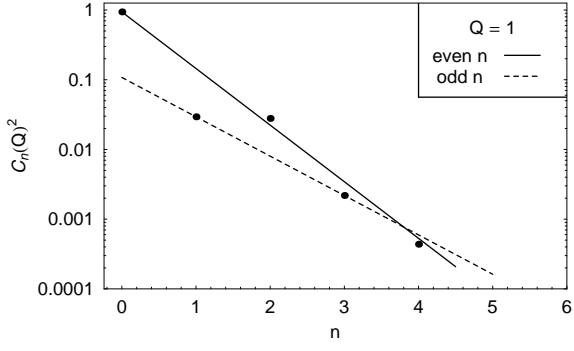}\hspace{2pc}%
\begin{minipage}[b]{18pc}
\caption{
Square of the expansion coefficients $C_n(Q)$ for $Q=1$. The ground state energy eigenvalue is $\varepsilon=1.1539\;\hbar\omega$. The coefficients of Eq. ($50$) are $f^{(e)}=1.87$, $f^{(o)}=1.30$.}
\end{minipage}
\end{figure}

\begin{figure}[t]
\begin{minipage}[b]{18pc}
\includegraphics[width=18pc]{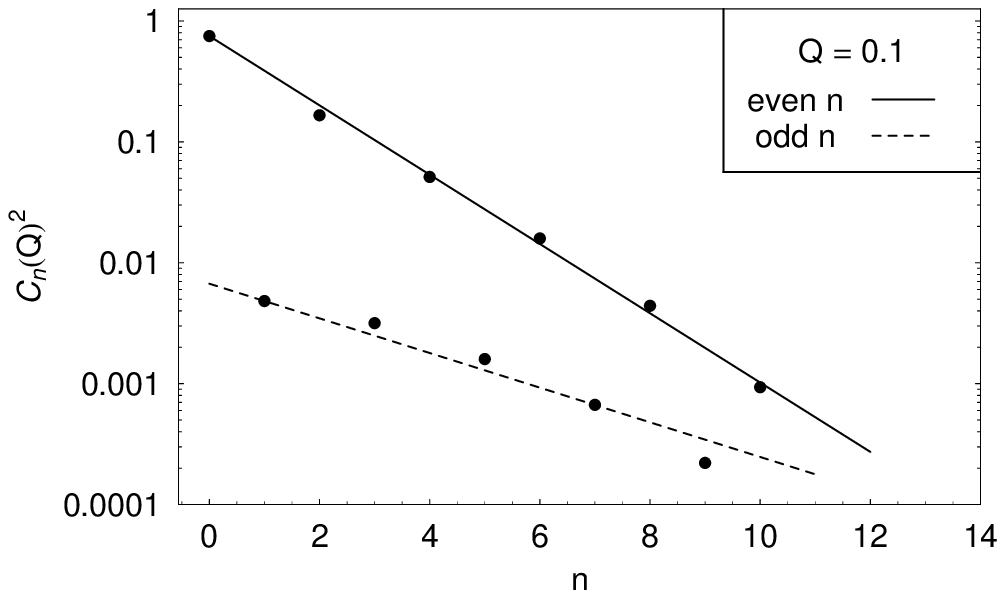}
\caption{Same as Fig. 4 for $Q=0.1$. $\varepsilon=2.7703\;\hbar\omega$. $f^{(e)}=0.66$, $f^{(o)}=0.33$.}
\end{minipage}\hspace{2pc}%
\begin{minipage}[b]{18pc}
\includegraphics[width=18pc]{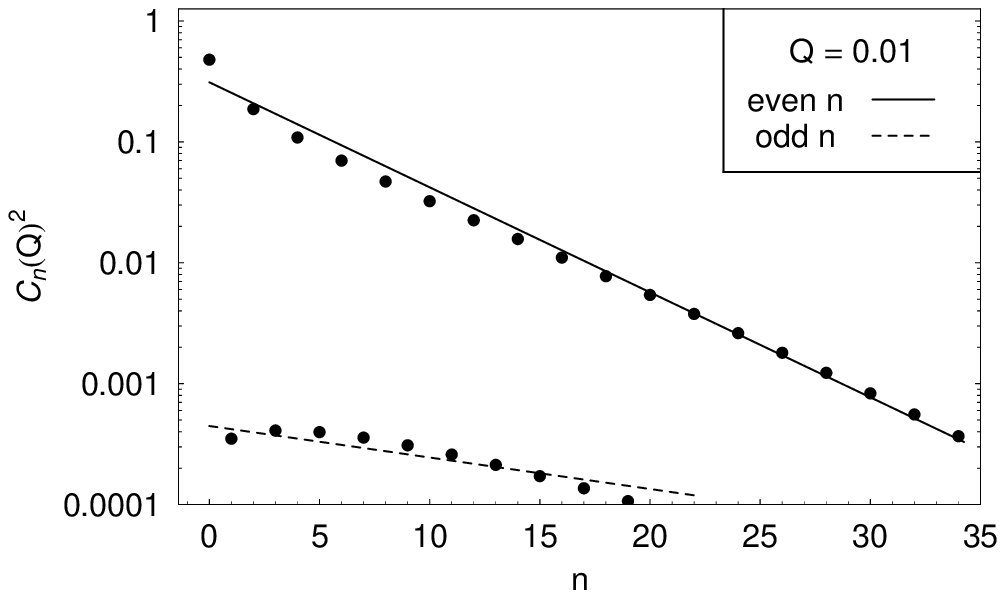}
\caption{Same as Fig. 4 for $Q=0.01$. $\varepsilon=8.4996\;\hbar\omega$. $f^{(e)}=0.2$, $f^{(o)}=0.06$.}
\end{minipage}
\end{figure}

\begin{equation}
\label{eq:50}
  C^{(e,o)}_{n}\propto \exp\left(-f^{(e,o)}(Q)\cdot n\right)\;,
\end{equation}
where $(e,o)$ stands for \emph{even} and \emph{odd} states, respectively. The values of $f^{(e/o)}(Q)$ are given in the captions. In Fig. 7, the Q-dependence of the ground state and the first and second excited state is shown.

\begin{figure}[t]
\begin{minipage}[b]{18pc}
\includegraphics[width=18pc]{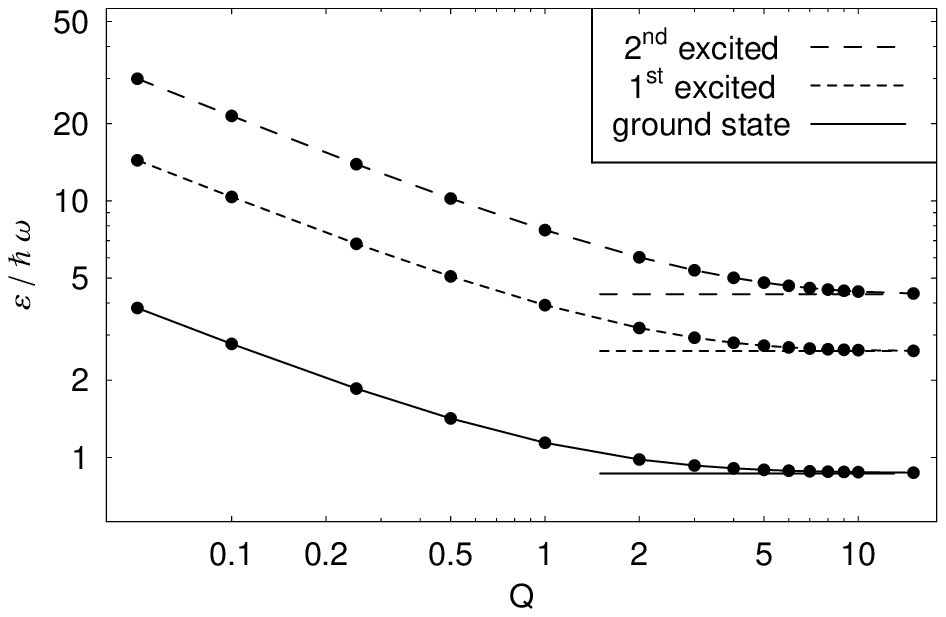}
\caption{Influence of the anharmonic effects on the excitation spectrum: $Q$-dependence of the three lowest energy eigenvalues corresponding to $\hat{H}(\bar{x}_2)$. The straight lines show the equidistant spectrum $\epsilon^{(2)}_{n}$ of Eq. (\ref{eq:27}), valid for $Q\rightarrow\infty$.}
\end{minipage}\hspace{2pc}%
\begin{minipage}[b]{19pc}
\includegraphics[width=18pc]{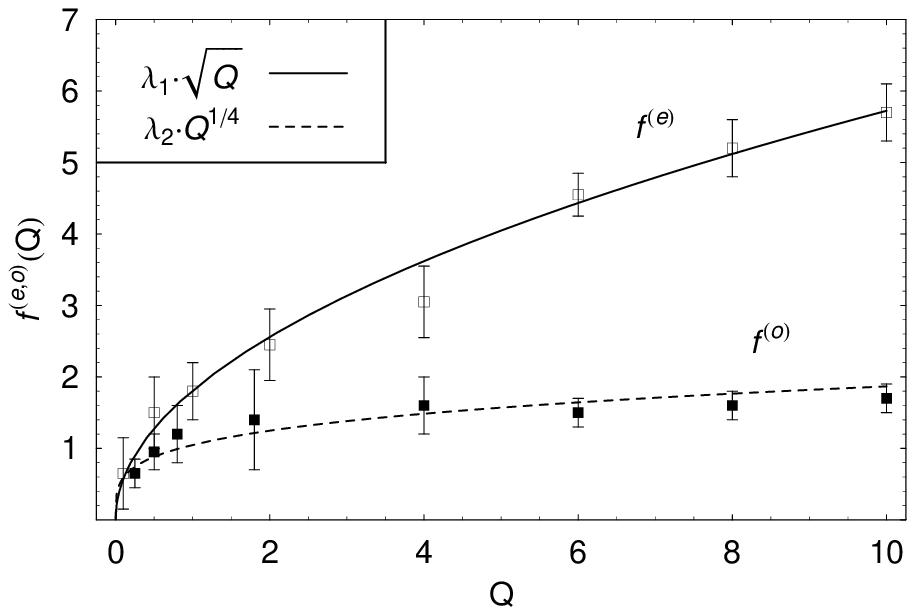}
\caption{
$Q$-dependence of the parameters $f^{(e,o)}$ in Eq. ($37$). The errorbars result from curve fitting with Eq. ($37$) and mark lines corresponding to minimum and maximum gradient, respectively. The lines are best fits with $\lambda_1=1.81$, $\lambda_2=1.05$.}
\end{minipage}
\end{figure}

In Fig. 8, the fit parameter $f^{(e,o)}(Q)$ in Eq. (\ref{eq:50}) is shown as a function of the coupling parameter. The Q-dependence is approximately given by: $f^{(e)}\propto\sqrt{Q}$ and $f^{(o)}\propto Q^{\frac{1}{4}}$. Thus, for $Q\leq 10$ the ground state wave function can be well approximated by
\begin{equation}
\label{eq:51}
  \frac{\Phi_0(\bar{x}_1,\bar{x}_2)}{\Phi_0(\bar{x}_1)}=\Phi_0(\bar{x}_2)=\sum_{n\,even}
  {\exp(-\lambda_1\, \sqrt{Q}\,n)\cdot|n\rangle}
  +\sum_{n\,odd}
  {\exp(-\lambda_2\, Q^{\frac{1}{4}}\,n)\cdot|n\rangle}\;,
\end{equation}
where $\Phi_0(\bar{x}_1)=\psi_0(\bar{x}_1)$ as in Eq. (\ref{eq:25}), and $\lambda_1$, $\lambda_2$ are given in the caption of Fig. 8.

\section{Discussion and Outlook}
In this paper, we presented analytical results for the wave function and energy spectrum of strongly correlated electrons. The results are exact for harmonic interaction as well as for arbitrary interaction in the strong coupling limit, $Q\rightarrow\infty$.

For quantum dots based on a \emph{GaAs}/\emph{AlGaAs} heterostructure (\mbox{$\epsilon_b=12.9$}, \mbox{$m=0.067\,m_e$}) with a characteristic intrinsic confinement energy of \mbox{$\hbar\,\omega=5$ meV}, we obtain the confinement frequency \mbox{$\omega=7.6$ THz} corresponding to a coupling parameter $Q=1.3$, an equilibrium distance (\ref{eq:14}) \mbox{$x_0=21.7$ nm}, and a width of the Gaussian peaks in the ground state electron density of Fig. 2 of \mbox{$\sqrt{\kappa}=1.3\,x_0$}. A situation of strong coupling of e.g. $Q=50$ would require a shallow confinement potential of \mbox{$\hbar\,\omega=0.088$ $\mu$eV} leading to \mbox{$x_0=32.1$ $\mu$m} and \mbox{$\sqrt{\kappa}=0.2\,x_0$}. Stronger coupling can be achieved in materials with strong Coulomb interaction. E.g. for a ZnSe structure ($\epsilon_b=9.1$, $m=0.21 m_e$) we obtain $Q=2.4$ (at $\hbar\,\omega=5$ meV).

For finite coupling, $Q\,\widetilde{>}\,1$, an analytical result is obtained by expanding the anharmonic corrections of the potential energy in terms of the $N$-particle eigenfunctions of the strong coupling (harmonic) limit. For moderate and small coupling, individual electrons overlap significantly. Therefore, in this case the resulting $N$-particle wave function has to be anti-symmetrized. The corresponding generalization of our method will be presented elsewhere. Furthermore, there are two aspects of our approach that will be examined in future work:

\emph{1. First principle spectrum from Quantum Monte Carlo data}. Path integral Monte Carlo (PIMC) simulations allow to compute the density matrix of the $N$-particle problem from first principles \cite{afilinov-etal.01prl,afilinov-etal.03pss}. However, they cannot directly yield the energy spectrum and corresponding wave functions. In the limit of strong correlations, the idea is to reconstruct the energy spectrum and wave functions from PIMC ground state results for the density distribution or correlation function using an expansion in terms of the derived oscillator eigenmodes
\begin{equation}
	\Psi_{0}(\vec{R})=\prod^{N\cdot d}_{i=1}{\left(\sum^{\infty}_{n=0}{C^{(i)}_{n}(Q)\cdot\psi_{n}(R_i) }\right)}\;,\hspace{1pc}E_0=V(\vec{r}_0)+\sum^{N\cdot d}_{i=1}{ \frac{\sum^{\infty}_{n=0}{C^{(i)}_n(Q)\,\hbar\Omega_i\left(n+\frac{1}{2}\right)}}{\sum^{\infty}_{n=0}{C^{(i)}_n(Q)}}}\;.
\end{equation}
E.g. a fit to the PIMC the density distribution
\begin{equation}
\rho_{QMC}(\vec{r})=\int {d^3 r_1\ldots d^3 r_N \,|\Psi_0(\vec{R})|^2\,\hat{\rho}(\vec{r})}\;,
\end{equation}
with $\vec{R}=\vec{R}(\vec{r}_1,\ldots,\vec{r}_N)$ and \mbox{$\hat{\rho}(\vec{r})=\sum^{N}_{i=1}{\delta(\vec{r}-\vec{r}_i)}$} then yields the expansion coefficients.

\emph{2. Application to Optics and Quantum Transport}. In the strong coupling limit, the $N$-particle Hamiltonian (\ref{eq:1}) including external fields, $\delta\hat{H}_{ext}$, can be written in second quantization as
\begin{equation}
\label{eq:a1}
	\hat{H}-V(\vec{r}_0) = \sum^{N\cdot d}_{i=1}{\hbar\,\Omega_i\left({\hat{a}}^{\dagger}_{i}\,{\hat{a}}_{i}+\frac{1}{2}\right)}
	+\,\delta \hat{V}\,+\,\delta\hat{H}_{ext}\;,
\end{equation}
where ${\hat{a}}^{\dagger}_{i}$, ${\hat{a}}_{i}$ are expected to be bosonic field operators of the derived eigenmodes satisfying the commutator relation
\begin{equation}
\label{eq:a2}
	\left[{\hat{a}}_{m},{\hat{a}}^{\dagger}_{n}\right]=\delta_{mn}\;.
\end{equation}
In terms of these field operators, one can define Nonequilibrium Green`s functions by 
\begin{equation}
\label{eq:a3}
	g^{<}_{m n}(t,t^{\prime})=\frac{1}{i\,\hbar}\langle\hat{a}^{\dagger}_{n}(t^{\prime})\,\hat{a}_{m}(t)\rangle\;,\hspace{2pc}
	g^{>}_{m n}(t,t^{\prime})=\frac{1}{i\,\hbar}\langle\hat{a}_{m}(t)\,\hat{a}^{\dagger}_{n}(t^{\prime})\rangle\;,
\end{equation}
and introduce density matrices
\begin{equation}
\label{eq:a5}
	F_{m n}(t)= i\,\hbar\,g^{<}_{m n}(t,t)\;.
\end{equation}
The subject of ongoing work is to derive equations of motion for $g^{<}_{n m}(t,t^{\prime})$ and $g^{>}_{n m}(t,t^{\prime})$ or $F_{m n}(t)$, respectively, the solutions of which give direct access to the optical and transport properties of the strongly correlated confined $N$-particle systems.

\appendix
\setcounter{section}{1}
\section*{Appendix. Explicit structure of $\delta V$ in Sec. 5}
$\delta V$ in Eq. ($37$) can be rewritten in terms of $y=\sqrt{\eta}\,\tilde{x}_2$ and $Q$ as follows ($i_{\max}=6$)
\begin{equation}
  \frac{\delta V}{\hbar \omega} = \sum^{6}_{i=3}{\frac{Q}{2^{\frac{1}{3}}}\left(-\frac{2^\frac{1}{6}}{\sqrt{\eta\; Q}}y\right)^{i}}=-\frac{2^{\frac{1}{6}}}{3^{\frac{3}{4}}\sqrt{Q}}y^3+\frac{2^{\frac{1}{3}}}{3 Q}y^4-\frac{\sqrt{2}}{3^{\frac{5}{4}} Q^{\frac{3}{2}}}y^5+\frac{2^{\frac{2}{3}}}{3 \sqrt{3} Q^2}y^6\,.
\end{equation}
Expanding $\delta V$ in (A.1) in terms of Hermite polynomials $H_k(y)$ yields
\[
  \frac{\delta V}{\hbar \omega} =\sum^{6}_{k=0}{\tilde{\nu}_k(Q) H_{k}(y)}\;,
\]
where the coefficients $\tilde{\nu}_k(Q)$ are given by
\begin{eqnarray}
\tilde{\nu}_0(Q)=\frac{5\cdot2^{\frac{2}{3}}\,{\sqrt{3}} + 6\cdot2^{\frac{1}{3}}\,Q }{24\,Q^2}\;&,&\hspace{1pc}
\tilde{\nu}_1(Q)=-\frac{5\cdot3^{\frac{3}{4}}\,{\sqrt{2}} + 6\cdot2^{\frac{1}{6}}\,3^{\frac{1}{4}}\,Q }
  {24\,Q^{\frac{3}{2}}}\;,\nonumber\\
\tilde{\nu}_2(Q)=\frac{5\cdot2^{\frac{2}{3}}\,{\sqrt{3}} + 4\cdot2^{\frac{1}{3}}\,Q }{16\,Q^2}\;&,&\hspace{1pc}
\tilde{\nu}_3(Q)=-\frac{5\,{\sqrt{2}}\,3^{\frac{3}{4}} + 3\cdot2^{\frac{1}{6}}\,3^{\frac{1}{4}}\,Q }
  {72\,Q^{\frac{3}{2}}}\;,\nonumber\\
\tilde{\nu}_4(Q)=\frac{5\cdot2^{\frac{2}{3}}\,{\sqrt{3}} + 2\cdot2^{\frac{1}{3}}\,Q }{96\,Q^2}\;&,&\hspace{1pc}
\tilde{\nu}_5(Q)=-\frac{1}{48\,{\sqrt{2}}\,3^{\frac{1}{4}}\,Q^{\frac{3}{2}}}\;,\hspace{2pc}
\tilde{\nu}_6(Q)=\frac{1}{96\,2^{\frac{1}{3}}\,{\sqrt{3}}\,Q^2}\;.\nonumber
\end{eqnarray}


\section*{References}
\begin{thebibliography}{99}
\bibitem{ashoori} R.C.~Ashoori, Nature {\bf 379}, 413 (1996)
\bibitem{reimann} S.M. Reimann, and M. Manninen, Rev. Mod. Phys. {\bf 74}, 1283 (2002)
\bibitem{afilinov-etal.01prl} A. Filinov, M. Bonitz, and Yu.E. Lozovik,
Phys. Rev. Lett. {\bf 86}, 3851 (2001)
\bibitem{ludwig-etal.03cpp} P. Ludwig, A. Filinov, M. Bonitz, and  Yu.E. Lozovik,
Contrib. Plasma Phys. {\bf 43}, 285 (2003)
\bibitem{afilinov-etal.03jpa} A. Filinov, M. Bonitz, and Yu.E. Lozovik,
J. Phys. A: Math. Gen. {\bf 36}, 5899 (2003)
\bibitem{peeters01} A. Matulis, and F.M. Peeters,
Solid State Comm. {\bf 117}, 655 (2001)
\bibitem{loz03} Yu.E. Lozovik, V.D. Mur, and N.B. Narozhny,
JETP {\bf 96}, 932 (2003)
\bibitem{loz04} Yu.E. Lozovik, V.D. Mur, N.B. Narozhny, and A.N. Petrosyan,
Laser Phys. Lett. {\bf 1}, No. 3, 154-160 (2004)
\bibitem{afilinov-etal.prb04} A. Filinov, C. Riva, F.M. Peeters, Yu.E. Lozovik, and
M. Bonitz, Phys. Rev. B {\bf 70}, 035323 (2004)
\bibitem{afilinov-etal.03pss} A. Filinov, P. Ludwig, V. Golubnychyi, M. Bonitz and Yu.E. Lozovik, phys. stat. sol. (c) 0, No. {\bf 5}, 1518-1522 (2003)
\end{thebibliography}
\end{document}